\documentclass[twocolumn,prl,aps,superscriptaddress,showpacs]{revtex4}
\usepackage{epsfig}

\begin{document}

\title{Existence of Orbital Order and its Fluctuation in Ba(Fe$_{1-x}$Co$_x$)$_2$As$_2$ Revealed by X-ray Absorption Spectroscopy}

\author{Y. K.  Kim}
\affiliation{Institute of Physics and Applied Physics, Yonsei University, Seoul 120-749, Korea}

\author{W. S. Jung}
\affiliation{Institute of Physics and Applied Physics, Yonsei University, Seoul 120-749, Korea}

\author{G. R. Han}
\affiliation{Institute of Physics and Applied Physics, Yonsei University, Seoul 120-749, Korea}

\author{K.-Y. Choi}
\affiliation{FPRD, Department of Physics and Astronomy, Seoul National University, Seoul 151-747, Korea}

\author{K.-H. Kim}
\affiliation{FPRD, Department of Physics and Astronomy, Seoul National University, Seoul 151-747, Korea}

\author{C.-C. Chen}
\affiliation{Stanford Institute for Materials and Energy Science, SLAC National Accelerator Laboratory, 2575 Sand Hill Road, Menlo Park, California 94025, USA}
\affiliation{Department of Physics, Stanford University, Stanford, California 94305, USA}

\author{T. P. Devereaux}
\affiliation{Stanford Institute for Materials and Energy Science, SLAC National Accelerator Laboratory, 2575 Sand Hill Road, Menlo Park, California 94025, USA}

\author{A. Chainani}
\affiliation{RIKEN SPring-8 Center, Sayo-cho, Hyogo, 679-5148, Japan}

\author{ J. Miyawaki}
\affiliation{RIKEN SPring-8 Center, Sayo-cho, Hyogo, 679-5148, Japan}

\author{Y. Takata}
\email[Deceased]{}
\affiliation{RIKEN SPring-8 Center, Sayo-cho, Hyogo, 679-5148, Japan}

\author{Y. Tanaka}
\affiliation{RIKEN SPring-8 Center, Sayo-cho, Hyogo, 679-5148, Japan}

\author{M. Ouara}
\affiliation{RIKEN SPring-8 Center, Sayo-cho, Hyogo, 679-5148, Japan}

\author{S. Shin}
\affiliation{RIKEN SPring-8 Center, Sayo-cho, Hyogo, 679-5148, Japan}
\affiliation{Institute for Solid State Physics, University of Tokyo, Kashiwa, Chiba 277-8581, Japan}

\author{A. P. Singh}
\affiliation{Pohang accelerator laboratory, Pohang University of Science and Technology, Pohang 790-784, Korea }

\author{ J.-Y. Kim}
\email[Electronic address:$~~$]{masson@postech.ac.kr}
\affiliation{Pohang accelerator laboratory, Pohang University of Science and Technology, Pohang 790-784, Korea }

\author{C. Kim}
\email[Electronic address:$~~$]{changyoung@yonsei.ac.kr}
\affiliation{Institute of Physics and Applied Physics, Yonsei University, Seoul 120-749, Korea}

\date{\today}

\begin{abstract}
We performed temperature dependent X-ray linear dichroism (XLD) experiments on an iron pnictide system, Ba(Fe$_{1-x}$Co$_x$)$_2$As$_2$ with $x$=0.00 and 0.05, to experimentally verify existence of orbital ordering (OO). We observed clear XLD in polarization dependent X-ray absorption spectra of Fe $L$ edges. By exploiting the difference in the temperature dependent behaviors, we were able to separate OO and structure contributions to XLD. The observed OO signal indicates different occupation numbers for $d_{yz}$ and $d_{zx}$ orbitals and supports existence of a ferro-OO. The results are also consistent with the theoretical prediction. Moreover, we find substantial OO signal above the transition temperature, which suggests that OO fluctuation exists well above the transition temperature.
\pacs{74.25.Jb,74.70.Xa,78.70.Dm}
\end{abstract}
\maketitle

%Introduction%%%

Various experimental results have shown that there are anomalous in-plane anisotropic behaviors in iron pnictides well below the transition temperature\cite{Dai,TMchuang}. A large anisotropy in spin flip energies $J_a$ and $J_b$ was obtained from inelastic neutron scattering experiments on CaFe$_2$As$_2$\cite{Dai}. Quasi particle interference patterns in scanning tunneling microscopy  result also suggest a large anisotropic behavior in electronic structure of CaFe$_2$As$_2$\cite{TMchuang}. However, such large anisotropic properties were not observed in other early experiments such as angle resolved photoemission spectroscopy (ARPES)\cite{Yi1,Liu,Feng}, X-ray absorption spectroscopy (XAS)\cite{Mannella} and optical measurements(IR)\cite{Wang}.

This inconsistency comes from the existence of twin domains\cite{Tanatar0}. However, it did not take long to devise a way to remove the twin domains\cite{Tanatar} and anisotropic behavior was indeed obtained in transport\cite{Tanatar,Fisher}, ARPES\cite{Kwan,Yi2,Shimo} and optical\cite{Masa} measurements. Observed anisotropy is not only quite large but also anomalous. Resistivity measurements reveal that the antiferromagnetic (AFM) direction is more conductive than the ferromagnetic (FM) direction\cite{Fisher}, which is opposite to the common expectation that AF ordering suppresses the electrical conductivity. This anomalous anisotropy cannot be explained by the structural anisotropy of the orthorhombic phase either. The difference in in-plane lattice parameters of the orthorhombic phase is less than 1 $\%$\cite{Simon} and such a small difference is not expected to explain the large anisotropy.

An additional effect is needed to explain such anisotropy and it was subsequently proposed that orbital ordering (OO) may play a key role\cite{Wei,Chi,Philip,Valen,Yong,Tom}. Observed anisotropy was between the AFM (often referred to as $a-$ or $x-$) and FM ($b-$ or $y-$) directions, which suggests that only $d_{yz}$ and $d_{zx}$ orbitals are important among the five Fe 3$d$ orbitals since other orbitals have R4 symmetry. In this respect, among the various possible orderings, ferro-OO which occurs through unequal occupation of $d_{yz}$ and $d_{zx}$ orbital states was propsed\cite{Chi,Valen,Yong,Tom}. The proposed OO is supposed to be responsible for the anisotropy in, for example, transport properties and also possibly drives the magnetic and structural transitions\cite{Wei}. It is therefore important to experimentally verify existence of such an OO.

It was proposed that temperature dependent X-ray linear dichroism (XLD) experiment may reveal the signature of OO\cite{Tom}. The difference between XAS data taken with linear polarizations along the AFM and FM directions (that is, XLD) is expected to have a specific spectral shape. To resolve the issue on the OO in pnictides, we performed XLD experiments on detwinned mother ($x$=0) and underdoped ($x$=0.05) Ba(Fe$_{1-x}$Co$_x$)$_2$As$_2$. Clear XLD signal was obtained from the results of both mother and underdoped compounds after the contribution from the structural anisotropy was removed.  Both the OO and structural signals were found to exist well above the transition temperature, indicating strong fluctuation.

\begin{figure}
\centering \epsfxsize=7.0cm \epsfbox{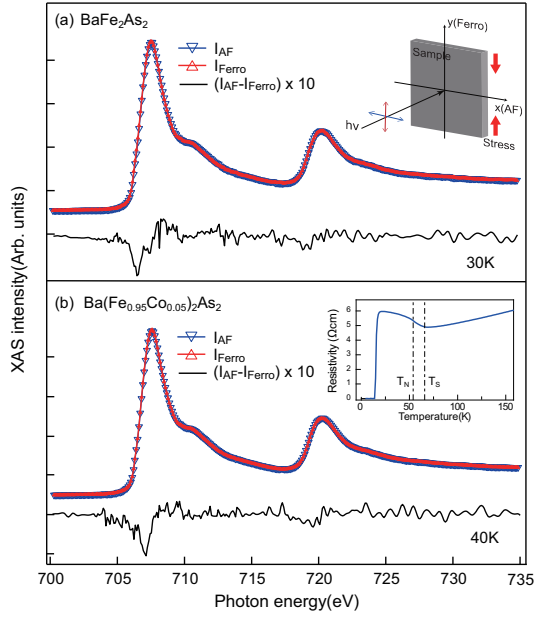} \caption{(Color online)(a) Fe $L_{3,2}$ edge XAS spectra from mother compound BaFe$_2$As$_2$ with two different polarizations E $\parallel$ $x$ (blue) and E $\parallel$ $y$ (red) at 30K. The lower curve is the XLD spectrum multiplied by 10 (black). The inset shows the experimental geometry and axes information. (b) Fe $L_2$ and $L_3$ edge XAS and XLD spectra for an underdoped compound Ba(Fe$_{1-x}$Co$_x$)$_2$As$_2$ taken at 40 K. The inset shows the temperature dependent resistivity. Resistivity was measured on a twinned crystal.} \label{fig1}
\end{figure}

\begin{figure}
\centering \epsfxsize=8.5cm \epsfbox{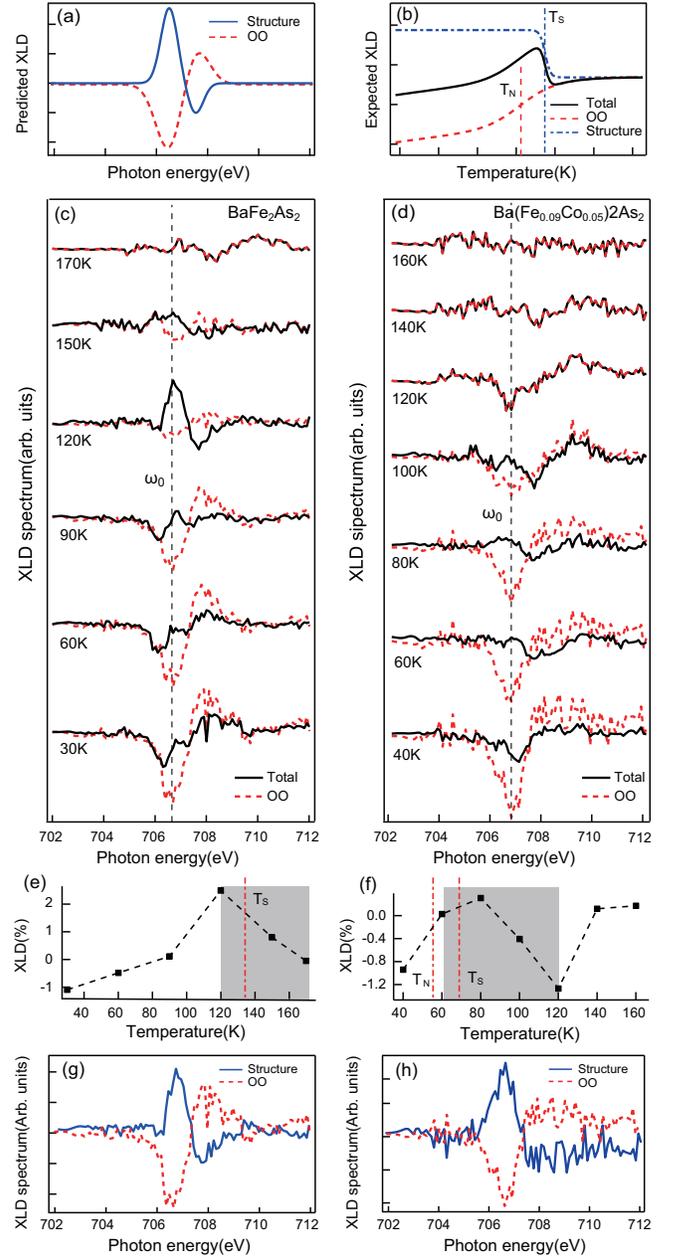} \caption{(Color online) (a) Schematic of the predicted OO and structure XLD signal\cite{Tom} and (b) their expected temperature dependence. Total XLD (solid black lines) and OO contribution (dashed red) at various temperatures for (c) mother and (d) underdoped compounds, respectively. Normalized XLD defined as (I$_{AFM}$-I$_{FM}$)/I$_{L_3}$ at $\omega_0$ as a function of temperature for (e) mother and (f) underdoped compounds. Structure (blue line) and OO (red line) contributions to XLD for (g) mother and (h) underdoped compounds.}
\label{fig2}
\end{figure}

%Experiment
Single crystals with two doping levels of $x$=0 and 0.05 were synthesized by the self-flux method\cite{SHKim}. Both structure ($T_S$) and magentic ($T_N$) transition temperatures of the mother compound were measured to be 135 K while underdoped sample showed $T_S$ and $T_N$ at 67 and 56 K, respectively (see the inset in Fig. 1(b)). Detwinning was achieved by applying mechanical stress on samples\cite{Kwan}. XAS experiments were performed at beam line 2A of Pohang Light Source (PLS) and at 17U of SPring-8. Samples were cleaved {\it in situ} under a pressure better than $1.0\times 10^{-7}$ Torr and were transferred immediately into the measurement chamber at a pressure better than $7.5\times 10^{-11}$ Torr. Spectra were taken at various temperatures. All the absorption spectra were recorded in the total electron yield mode and were normalized by the incident photon flux at a gold mesh. The direction of light polarization was controlled by rotating the sample.

%Results : Fig1 Absorption spectrums and observation of XLD signal
Figure 1 shows Fe $L_{3,2}$ edge XAS spectra taken with polarizations along the FM ($y-$) and AFM ($x-$) directions as well as XLD signal (difference) for (a) mother and (b) underdoped compounds. The experimental geometry as well as the definition of the axes are shown in the inset of Fig. 1(a). Axes can be defined with respect to the direction of applied stress. To obtain the XLD signal, all XAS spectra were normalized using the following procedure. First, linear background was removed to make the tail part parallel to the energy axis. The spectra were then normalized by the area normalization method. After these steps, we could obtain the XLD spectrum by a simple subtraction of two normalized spectra. The resulting spectra show small but clear XLD signal. With linear polarizations along the $x-$ and $y-$axes, the only possible dichroism signal comes from the difference between $d_{yz}$ and $d_{zx}$ orbital occupations because others have the same parity with respect to both $xz-$  and $yz-$planes. Therefore, the observed XLD implies that occupations of $d_{yz}$ and $d_{zx}$ orbital states are indeed different.

%Results : Fig2 Structure effect consideration

%Figure 2(a) and 2(b)
However, one has to be careful in interpreting the XLD signal in Fig. 1 as being due to OO because it also contains contributions from the structural anisotropy. The system has an orthorhombic structure with the broken R4 symmetry below $T_S$, and thus we may expect additional XLD purely from symmetry reason without an OO. Especially, the predicted structural contribution has a behavior opposite to that from OO (see Fig. 2(a)), which may make the XLD signal small.\cite{Tom} Therefore, we need to find a way to separate the two contributions in the XLD. We argue that the structure contribution has an abrupt change across the $T_S$ and saturates right away unlike the OO contribution which is aspected to slowly enhances (see Fig. 2(b))\cite{Chen}. The abrupt change for the structure contribution is justified from the fact that lattice constants also abruptly change at the $T_S$ and do not show appreciable change well below the $T_S$ even with an external stress\cite{Rotter,Blomberg}. One can exploit such different temperature dependences of OO and structure contributions to separate out the OO contribution as detailed below.

%Figure 2 (c), (d), (e), (f)
We performed temperature dependent XLD experiment and the results are plotted in Figs. 2(c) and (d). The total XLD (black lines) has a clear temperature dependent behavior. Upon a closer inspection, the behavior is rather anomalous. Instead of monotonic decrease with temperature lowering, the XLD has a pronounced behavior near $T_S$. In addition, the shape of XLD spectrum changes dramatically around $T_S$. Particularly, for the case of mother compound, the spectrum at 120 K not only shows more pronounced dichroism but also has opposite behavior near the edge compared to the 30 K spectrum. Such pronounced XLD signal around $T_S$ is mostly from the structure contribution because OO contribution has not yet set in. As the temperature is lowered even further, OO contribution finally comes in and somewhat cancels out the structure contribution, resulting in a smaller overall XLD signal.

However, actual structure contribution does not appear abruptly at $T_S$ as discussed above. To trace the detailed temperature dependence of structure contribution, we plot the magnitude of XLD signal at $\omega_0$ divided by $L_3$ main peak intensity as a function of temperature in Figs. 2(e) and (f) for mother and underdoped compounds, respectively. In both plots, we see that XLD signal increases within a certain temperature range around $T_S$ (indicated by gray area) as the temperature is lowered. This increase is due to appearance of the structure contribution. Note that the temperature at which the XLD signal starts to increase is higher than $T_S$. This implies that the structure transition also occurs from that temperature. The structure contribution finally saturates at a temperature below $T_S$ where the XLD signal decrease again. Such broadening of the structure transition temperature could be attributed to the effect from the applied external stress which was observed in recent studies\cite{Blomberg,Chetan}

\begin{figure}
\centering \epsfxsize=8.5 cm \epsfbox{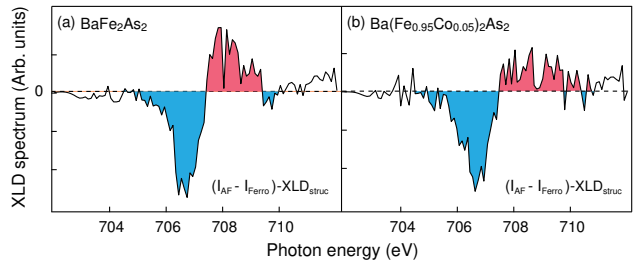} \caption{(Color online) OO contribution to the XLD signal for (a) mother compound and (b) underdoped compound.}
\label{fig3}
\end{figure}

We also noted differences in the temperature dependent XLD from the two compounds. In the underdoped case, XLD shows an down-turn behavior (between 120 and 140 K) before it starts to increase while it is not seen in mother compound case. In addition, XLD in underdoped is significantly stronger than that in mother compound. These facts imply that OO in underdoped compound starts to appear at a much higher temperature than $T_S$ and than in mother compound. XLD spectrum of underdoped compound taken at 120 K also shows small dichroism signal that is consistent with the shape of OO contribution. Detailed discussion on this doping dependence will be given later.

With this difference in the temperature dependence, we successfully separated the OO and structure contribution from the total XLD. First, to extract structure contribution, we subtract the spectra taken above T$_S$ (150 K for mother and 120 K for underdoped compounds) from that taken just below the transition temperature (120 K for mother and 80 K for underdoped compounds). We must mention that there is a finite change in the OO contribution between the two temperatures which must be considered in estimating the true structure contribution. It is reflected in our data analysis but the details will not be discussed here. The resulting spectrum is the structure signal and is plotted in Figs. 2(g) and (h) (blue lines). Shapes for different compositions are similar and are consistent with the theoretical prediction\cite{Tom}. Then, the OO contribution can be easily obtained by subtracting the structural contribution from the total XLD signal based on the decreasing behavior mentioned before. Resulting OO contribution is also plotted in Figs. 2 (g) and (h) as red dashed lines. The behavior is again quite consistent with the predicted one schematically shown in Fig. 2(a)\cite{Tom}. OO contributions at other temperatures are overlaid in Figs. 2(c) and 2(d) (red dashed lines). Note that all the OO spectra have a consistent shape and only the magnitude decreases as the temperature increases. Eventually, the OO contribution disappears at around 170 K for mother and 140 K for underdoped compounds. However, 170 K spectrum from mother compound shows a small dichroism signal but we attribute it to signal due to out-of-plane polarization as observed earlier\cite{Mannella}.

%%%%Fig.3 Analysis of signal from orbital ordering
We now analyze the OO contribution in detail. We re-plot in Fig. 3 the data in Figs. 2(g) and 2(h). Both spectra show a dip feature at photon energy of 706.7 eV. On the other hand, a peak feature appears at different energies; a peak at 708 eV for the mother compound and a broad feature at 709.3 eV for the underdoped system. As we aligned the light polarization along the $x-$ (AFM) and $y-$ (FM) directions , spectrum taken with the $x-$ ($y-$) polarization detects $d_{zx}$ ($d_{yz}$) orbital.  Therefore, a dip in the data means $d_{yz}$ has higher density of states than $d_{zx}$. This fact directly connects to unequal occupation numbers for $d_{yz}$ and $d_{zx}$, $\it {i.e.}$ $d_{yz}$ state is less occupied than $d_{zx}$ as theoretically predicted\cite{Chi,Valen,Yong,Tom}. On the other hand, a peak feature on the higher energy side implies the opposite situation. However, we note that the area of the dip feature is larger than that of the peak feature for both of the compounds, which is again consistent with the theoretical prediction. Theoretically predicted XLD magnitude is around 3 $\%$ while our experimental values are 4.25 $\%$ and 3.74 $\%$ for mother and underdoped compounds, respectively.

\begin{figure}
\centering \epsfxsize=8.5 cm \epsfbox{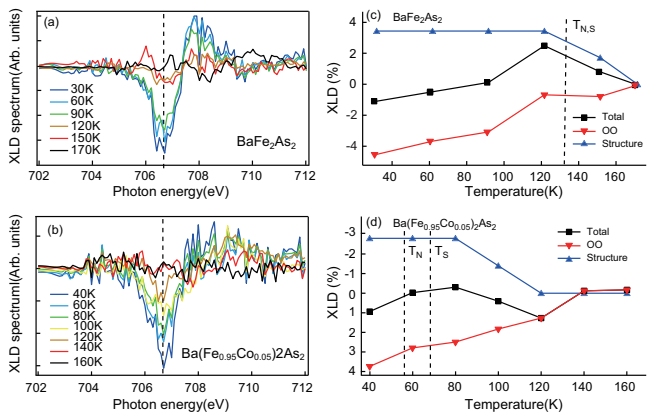} \caption{(Color online)  Overlaid OO contributions at various temperatures for (a) mother and (b) underdoped compounds. Temperature dependence of the XLD magnitude for total (black), OO (red) and structure (blue) for (c) mother and (d) underpdode compounds.}
\label{fig4}
\end{figure}

Before we move onto discussing details of the temperature dependence, we wish to briefly touch upon possible contribution of magnetic origin to XLD. We believe magnetic XLD is negligible considering the fact that there is an excellent agreement between theoretical predictions and experimental results of OO signal. This view is also consistent with an earlier assertion that XLD signal from Ni$^{2+}$ is simply the property of wave functions of $d$-orbital states, which is drawn based on the comparison of experiments and atomic multiplet calculations\cite{Elke}.

%Fig.4 Detailed T dependence of OO signal and fluctuation above the transition temperature
For a detailed investigation of the temperature dependence, we plot OO contribution again in Figs. 4(a) and 4(b). A pronounced and monotonic temperature dependence is already seen. In Figs. 4(c) and 4(d), we plot the magnitudes of the structure and OO contributions as a function of temperature along with the total XLD. Unlike the total XLD, partial contributions show monotonic behaviors for both compositions. A noticeable aspect of the data is that both structure and OO contributions appear well above the transition temperature. For the structure contribution, its presence above the $T_S$ could be attributed to an extrinsic effect due to the external stress for detwinning.\cite{Blomberg,Chetan}

Turning our attention to the OO contribution, its appearance above the transition temperatures indicates existence of strong OO fluctuation above the transition temperature. It is consistent with recent resistivity result\cite{Fisher} and anisotropic band shifts observed in ARPES data\cite{Yi2}. In both results anisotropic behaviors appear above the transition temperatures around 150 K for underdoped samples. Such anisotropy above the transition temperatures was attributed to the effect of applied stress which makes spin fluctuation stabilized and pushes $T_N$ to a higher temperature\cite{Chetan}. Co-existence and a similar behavior of OO and spin fluctuations may suggest a close relation between the two. A possibility is that stabilization of OO by the stress drives spin ordering through spin orbit coupling.

On the other hand, the temperature range in which OO fluctuation exists increases with doping. It is about 45 K for mother compound and 80 K for underdoped case. This indicates enhancement of the fluctuation effect upon doping. The presence of OO fluctuation as well as its enhancement upon doping could mean an important role of OO fluctuation for the superconductivity in iron pnictides. Indeed, it was recently proposed that OO fluctuation could act as a glue for the pairing in iron pnicitides. An important aspect of OO fluctuation mediated superconductivity is that the superconducting gap has s$_{++}$ symmetry\cite{Kontani,Shimo2}. The gap symmetry, whether it is s$_{++}$ or s$_{+-}$, is a crucial information in understanding the superconductivity. Our finding supports the discussion in a recent report\cite{Kontani} and could explain why some systems seem to have s$_{++}$ gap symmetry while others s$_{+-}$, depending on which of OO or spin fluctuation prevails. A comparative XLD studies of materials with s$_{++}$ or s$_{+-}$ gap symmetries could further resolve the issue.

%Acknowledgments%%
\acknowledgments
Authors would like to thank Ch. Kim and K. D. Lee for experimental supports. This work was supported through NRF Grant No. 20100018092 and KICOS Grant No. K20062000008. The work at SNU was supported by the National Creative Research Initiative Grant No. 2010-0018300. The work at Spring-8 was supported by Grants-in-Aid for Scientific Research (A) Grant No. 21244049 from JSPS and was performed with the approval of RIKEN (Proposal No. 20110028). PLS is supported by BSRP through the NRF funded by the MEST (2009-0088969).


\begin{thebibliography}{28}


\bibitem{Dai} Jun Zhao \emph{et al}, Nat. Phys. {\bf 5}, 555 (2009).

\bibitem{TMchuang} T.-M. Chuang \emph{et al}, Science {\bf 327}, 171 (2010).

\bibitem{Yi1} M. Yi \emph{et al}, Phys. Rev. B {\bf 80}, 174510 (2009).

\bibitem{Liu} Chang Liu \emph{et al}, Phy. Rev. Latt. {\bf 102} 167004 (2009).

\bibitem{Feng} L. X. Yang \emph{et al}, Phys. Rev. Lett. {\bf 102}, 107002 (2009).

\bibitem{Mannella} C. Parks Cheney \emph{et al}, Phys. Rev. B {\bf 81}, 104518 (2010).

\bibitem{Wang} W. Z. Hu \emph{et al}, Phys. Rev. Lett. {\bf 101}, 257005 (2008).

\bibitem{Tanatar0} M. A. Tanatar \emph{et al}, Phys. Rev. B {\bf 79}, 180508 (2010).

\bibitem{Tanatar} M. A. Tanatar \emph{et al}, Phys. Rev. B {\bf 81}, 184508 (2010).

\bibitem{Fisher} Jiun-Haw Chu \emph{et al}, Science {\bf 329}, 824 (2010).

\bibitem{Kwan} Yeongkwan Kim \emph{et al}, Phys. Rev. B {\bf 83}, 064509 (2011).

\bibitem{Yi2} M. Yi \emph{et al}, Proc. Natl. Acad. Sci. USA {\bf 108}, 6878 (2011).

\bibitem{Shimo} T. Shimojima \emph{et al}, Phys. Rev. Lett. {\bf 104}, 057002 (2010).

\bibitem{Masa} M. Nakajima \emph{et al}, Proc. Natl. Acad. Sci. USA {\bf 108}, 12238 (2011).

\bibitem{Simon} Simon A. J. Kimber \emph{et al}, Nat. mat. {\bf 8}, 471 (2009).

\bibitem{Wei} Weicheng Lv, Jiansheng Wu, and Philip Phillips, Phys. Rev. B {\bf 80}, 224506(2009).

\bibitem{Chi} Chi-Cheng Lee, Wei-Guo Yin, and Wei Ku, Phys. Rev. Lett. {\bf 103}, 267001 (2009).

\bibitem{Philip} Weicheng Lv, Frank Kr\"{u}ger, and Philip Philips, Phys. Rev. B {\bf 82}, 045125(2010).

\bibitem{Valen} B. Valenzuela, E. Bascones, and M. J. Calder$\acute{O}$n, Phys. Rev. Lett. {\bf 105}, 207202 (2010).

\bibitem{Yong} Da-Yong Liu \emph{et al}, Phys. Rev. B {\bf 84}, 064435(2011).

\bibitem{Tom} C.-C. Chen \emph{et al}, Phys. Rev. B {\bf 82}, 100504(R) (2010).

\bibitem{SHKim} S. H. Kim \emph{et al}, J. Appl.  Phys. {\bf 108}, 063916 (2010).

\bibitem{Chen} C.-C. Chen, \emph{et al}, Phys. Rev. B {\bf 80}, 180418(R) (2009).

\bibitem{Rotter} Marianne Rotter \emph{et al}, Phys. Rev. B {\bf 78}, 020503(R) (2008).

\bibitem{Blomberg} E. C. Blomberg \emph{et al}, arXiv:1111.0997v1 (2011).

\bibitem{Chetan} Chetan Dhital \emph{et al}, arXiv:1111.2326v1 (2011).

\bibitem{Elke} Elke Arenholz \emph{et al}, Phys. Rev. Lett. {\bf 98} 197201 (2007).

\bibitem{Kontani} Tesuro Saito, Seiichiro Onari, Hiroshi Kontani, Phys. Rev. B {\bf 82}, 144510 (2010).

\bibitem{Shimo2} T. Shimojima \emph{et al}, Science {\bf 332}, 564 (2011).

%\bibitem{He} C. He, Y. Zhang, X. F. Wang, L. X. Yang, B. Zhou, F. Chen, M. Arita, K. Shimada, H. Namatame, M. Taniguchi, X. H. Chen, J. P. Hu and D. L. Feng, Phys. Rev. Lett. {\bf 105}, 117002 (2010).

\end{thebibliography}
\end{document}